\documentclass[aps,twocolumn]{revtex4}
\usepackage{amsfonts}

\usepackage{amsmath}

\input{tcilatex}

\begin{document}

\title{Minimal length in Quantum Mechanics and non-Hermitian Hamiltonian systems}
\author{Bijan Bagchi$^\circ$ and Andreas Fring$^\bullet$}

\begin{abstract}
Deformations of the canonical commutation relations lead to non-Hermitian
momentum and position operators and therefore almost inevitably to
non-Hermitian Hamiltonians. We demonstrate that such type of deformed
quantum mechanical systems may be treated in a similar framework as
quasi/pseudo and/or $\mathcal{PT}$-symmetric systems, which have recently
attracted much attention. For a newly proposed deformation of exponential
type we compute the minimal uncertainty and minimal length, which are
essential in almost all approaches to quantum gravity.
\end{abstract}

\preprint{HEP/123-qed}

\affiliation{$\circ$ Department of Applied Mathematics, University of Calcutta 92 Acharya
Prafulla Chandra Road, Kolkata 700 009, India\\
$\bullet$ Centre for Mathematical Science, City University London,
Northampton Square, London EC1V 0HB, UK\\
E-mail: BBagchi123@rediffmail.com, A.Fring@city.ac.uk}

\maketitle


\section{Introduction}

The conventional space-time structure described locally in flat Minkowski
space has not been confirmed experimentally up to several orders of the
Planck length. This fact allows for the possibility to accommodate various
modifications of the short distance structure which are needed to achieve
consistency in several types of quantum theories, especially those which aim
to incorporate gravity \cite{calmet}. Numerous investigations in string
theory \cite{String1} and alternative approaches to quantum gravity \cite
{Ashtekar} have indicated the necessity to introduce a so-called minimal
length, which constitutes a bound beyond which the localization of
space-time events is no longer possible. Such type of limitations inevitably
require some generalizations of the uncertainty relations, which usually
originate from a modification of the related canonical commutation relations 
\cite{Kempf2,Brodimas,Wess1,Bieden,QuesneTK}. There exist also alternative
mechanisms giving rise to a minimal length, but ultimately these approaches
are all related \cite{Hossenfelder}. Modifying the canonical commutation
relations will almost unavoidably lead to operators, which are non-Hermitian
or strictly speaking not self-adjoint and therefore they are not observable.
Consequently the Hilbert space structure needs to be modified. This is a
familiar scenario encountered in the context of non-Hermitian Hamiltonian
systems with real eigenvalues; a field of interest initiated around ten
years ago by the seminal paper \cite{Bender:1998ke} and which is a topic
currently also explored experimentally \cite{Expp}, see \cite{Benderrev} for
a recent review. However, in the latter context the starting point is very
different, namely a Hamiltonian rather than the set of canonical variables.
The main objective here is to explore the similarities between these two
scenarios and unravel their differences.

\section{Quasi/Pseudo Hermitian versus deformed Quantum Mechanics}

Non-Hermitian Hamiltonian systems with real eigenvalues allow for a
consistent quantum mechanical description, may they be either
quasi-Hermitian \cite{Dieu}, pseudo-Hermitian \cite{pseudo1} or/and $%
\mathcal{PT}$-symmetric \cite{Bender:2002vv}. When dealing with such type of
systems one usually starts from some physical or mathematical motivation for
a non-Hermitian Hamiltonian $H\neq H^{\dagger }$. Next one seeks a positive
operator $\rho $, whose adjoint action corresponds to the Hermitian
conjugation $\rho H\rho ^{-1}=H^{\dagger }$. Factorizing this operator into
a product of a Dyson operator $\eta $ and its Hermitian conjugate in the
form $\rho =\eta ^{\dagger }\eta $ allows in a sufficient manner to compute
an isospectral Hermitian Hamiltonian counterpart from $h=\eta H\eta
^{-1}=h^{\dagger }$. A well known and important feature in this context is
the fact that when given only the Hamiltonian $H$ the subsequently
constructed operator $\rho ,$ and therefore also $\eta $ and $h$, is not
unique. However, as was argued in \cite{Dieu} uniqueness can be achieved by
one additional choice: One could fix directly the transforming operators $%
\rho $, $\eta $ or one physical observable $o^{\prime }$ or $\mathcal{O}%
^{\prime }$ either in the Hermitian or non-Hermitian system, respectively.
These two choices, i.e. a definite form for the Hamiltonian $H$ and an
additional observable, will fix the metric uniquely, such that there are no
ambiguities left in the interpretation of the physical observables. In the
corresponding Hermitian Hamiltonian counterpart description of the system
all physical observables $o$ may be determined from $\eta \mathcal{O}\eta
^{-1}=o$.

Schematically summarized the above can be described by the following
sequence of steps
\begin{equation}
\begin{array}{r}
H\neq H^{\dagger }\overset{\rho }{\Rightarrow }\rho H\rho ^{-1}=H^{\dagger }~%
\overset{\rho =\eta ^{\dagger }\eta }{\Rightarrow }~\eta H\eta
^{-1}=h=h^{\dagger } \\ 
\overset{o^{\prime },\mathcal{O}^{\prime },\eta ^{\prime }}{\Rightarrow }%
\quad \eta \mathcal{O}\eta ^{-1}=o.~~~\ 
\end{array}
\label{HH}
\end{equation}
As a consequence of the non-Hermiticity of $H$ its eigenstates no longer
form an orthonormal basis and the Hilbert space representation has to be
modified. This is achieved by utilizing the operator $\rho $ as a metric to
define a new inner product $\langle \quad |\quad \rangle _{\rho }$ in terms
of the standard inner product $\langle \quad |\quad \rangle $ as $\langle
\Phi |\Psi \rangle _{\rho }:=\langle \Phi |\rho \Psi \rangle $, for
arbitrary states $\langle \Phi |$ and $|\Psi \rangle $. Observables $%
\mathcal{O}$ are then by construction Hermitian with regard to this metric $%
\langle \Phi |\mathcal{O}\Psi \rangle _{\rho }:=\langle \mathcal{O}\Phi
|\Psi \rangle _{\rho }$.

The treatment of deformed quantum mechanical systems is somewhat similar,
although there are some crucial differences regarding the uniqueness of the
construction resulting from a contrary starting point. In this latter
context one commences from a modified version of the commutation relation
between the dynamical variables corresponding to the position operator $X$
and momentum operator $P$ 
\begin{equation}
\lbrack X(\alpha ),P(\alpha )]=i\hbar f(X(\alpha ),P(\alpha )),  \label{non}
\end{equation}
for some arbitrary function $f$ and $\alpha $ being a deformation parameter
or possibly a collection of them. One recovers Heisenberg's canonical
commutation relations for the standard momentum $p_{0}$ and position
operator $x_{0}$ in some well defined limit
\begin{equation}
\lim\nolimits_{\alpha \rightarrow 0}(X(\alpha ),P(\alpha ))=(x_{0},p_{0})~~%
\text{with }[x_{0},p_{0}]=i\hbar .  \label{00}
\end{equation}
In general, the operators defined in (\ref{non}) are non-Hermitian, i.e. $%
X^{\dagger }\neq X$, $P^{\dagger }\neq P$, and one therefore has to proceed
similarly as indicated in (\ref{HH}). First one seeks the metric operator $%
\rho $ such that its adjoint action Hermitian conjugates the new variables $X
$ and $P$ as $\rho (X,P)\rho ^{-1}=(X^{\dagger },P^{\dagger })$. As opposed
to the treatment of non-Hermitian Hamiltonian systems with real eigenvalues
this metric will be unique, because one has already selected two observables
from the very beginning which is sufficient according to the argumentation
in \cite{Dieu}. Next one has to select a Hamiltonian in order to specify the
physical system one wishes to describe. This is most naturally formulated in
terms of the variables $X$ and $P$, such that non-Hermitian Hamiltonians,
i.e. $H^{\dagger }(X,P)\neq H(X,P)$, almost inevitably arise from
deformations of the type (\ref{non}), when the undeformed Hamiltonian $%
H(x_{0},p_{0})=\lim_{\alpha \rightarrow 0}H(X,P)$ was taken to be Hermitian.
The deformed quantum mechanical system is now uniquely determined. The
Hilbert space is constructed in the same manner as above, that is by a
re-definition of the inner product by utilizing the metric operator $\rho $.
Factorizing the metric operator into a product of $\eta $ and its Hermitian
conjugate allows as an alternative view to consider the entire system in the
equivalent picture of its isospectral Hermitian counterpart
\begin{equation}
\eta (X,P,H,\ldots )\eta ^{-1}=(x,p,h,\ldots )=(x^{\dagger },p^{\dagger
},h^{\dagger },\ldots ).  \label{4}
\end{equation}
When the undeformed Hamiltonian $H(x_{0},p_{0})$ is not Hermitian one also
has to carry out the steps in (\ref{HH}) as $X$ and $P$ are no longer
observables in that case.

The procedure works as shown below 
\begin{equation}
\begin{array}{l}
(x_{0},p_{0})\!\rightarrow \!\!(X,P)\overset{\rho }{\Rightarrow \!}\rho
(X,P)\rho ^{-1}\!=\!(X^{\dagger },P^{\dagger })\overset{\eta }{\!\Rightarrow 
}\!\!(x,p) \\ 
=\eta (X,P)\eta ^{-1}\Rightarrow \eta H(X,P)\eta ^{-1}\!=\!\left\{ 
\begin{array}{l}
\!\!h(x,p) \\ 
\!\!H(x,p)\Rightarrow (\ref{HH})
\end{array}
\right.
\end{array}
\label{sum}
\end{equation}
We illustrate now the above general statements with some concrete examples:

There are various possibilities to deform the standard canonical commutation
relations between the dynamical variables $X$ and $P$. One might for
instance deform Heisenberg's canonical commutation relations as $%
PX-qXP=i\hbar $ \cite{Wess1}. Here we will instead assume a $q$-deformation
of the corresponding commutation relations between the creation operator $%
a^{\dagger }$ and the annihilation operator $a$ in the general form 
\begin{equation}
aa^{\dagger }-q^{2}a^{\dagger }a=q^{g(N)},~\qquad \text{with }N=a^{\dagger }a
\label{aa}
\end{equation}
where $g$ is some arbitrary function of the number operator $N$. The case $g$
taken to be just $0$ corresponds to the deformation studied for example in 
\cite{Kempf2}, whereas when $g(N)=N$ we recover the version explored for
instance in \cite{Brodimas}. In both cases the deformed Fock space can be
constructed explicitly \cite{Bieden}.

Assuming the representation for $X$ and $P$ to be still linear in $a$ and $%
a^{\dagger }$ we define 
\begin{equation}
X=\alpha a^{\dagger }+\beta a,~~~P=i\gamma a^{\dagger }-i\delta a,\qquad ~%
\text{ }\alpha ,\beta ,\gamma ,\delta \in \mathbb{R}.
\end{equation}
Then with the help of (\ref{aa}) we can write $[X,P]~$as
\begin{eqnarray}
\!\!\!\!\! &&\!\!\!\!\!\!\!\left[ X,P\right] =i\hbar q^{g(N\ )}(\alpha
\delta +\beta \gamma )  \label{gxp} \\
&&+\frac{i\hbar (q^{2}-1)}{\alpha \delta +\beta \gamma }\left( \delta \gamma
X^{2}+\alpha \beta ~P^{2}+i\alpha \delta XP-i\beta \gamma PX\right) ,  \notag
\end{eqnarray}
together with the constraint $4\alpha \gamma =(q^{2}+1)$. The canonical
commutation relations (\ref{00}) are obtained in the simultaneous limit $%
(\alpha \delta +\beta \gamma )\rightarrow 1$, $q\rightarrow 1$. The
relations (\ref{gxp}) simplify by taking the limit $\beta \rightarrow \alpha 
$, $\delta \rightarrow \gamma $, in which case the commutator (\ref{gxp})
reduces to a version studied for instance in \cite{Kempf2} for the special
case $g(N)=0$ 
\begin{equation}
\left[ X,P\right] =i\hbar q^{g(N)}+\frac{i\hbar }{4}(q^{2}-1)\left( \frac{%
X^{2}}{\alpha ^{2}}+\frac{P^{2}}{\gamma ^{2}}\right) .  \label{xp}
\end{equation}
We may simplify (\ref{xp}) further by taking different limits to obtain a
pure $P$-dependence on the right hand side.

Taking $g(N)=0$ in (\ref{xp}), parameterizing the deformation parameter in
the form $q=e^{2\tau \gamma ^{2}}$ with $\tau \in \mathbb{R}^{+}$ and taking
the limit $\gamma \rightarrow 0$ we obtain the simple deformation
\begin{equation}
\left[ X,P\right] =i\hbar \left( 1+\tau P^{2}\right) .  \label{12}
\end{equation}
It is easy to find a representation for $X$ and $P$ which will reproduce (%
\ref{12}) in terms of the standard momentum and position operators $p_{0}$
and $x_{0}$, respectively. We may for instance select $X=(1+\tau
p_{0}^{2})x_{0}$ and $P=p_{0}$. The undeformed case is obviously recovered
in the limit $\tau \rightarrow 0$. As announced in the previous section we
find that the operators associated to the deformed commutation relations are
in general not Hermitian $X^{\dagger }=X+2\tau i\hbar P~$and $P^{\dagger }=P$%
, albeit the simplified version (\ref{12}) still allows one operator to
remain Hermitian. According to (\ref{sum}), the unique metric operator $\rho 
$ is constructed to $\rho =(1+\tau P^{2})^{-1}$. At this point the
positivity of $\tau $ becomes important, as it ensures the absence of
singularities in the metric. In a momentum space representation $%
x_{0}=i\hbar \partial _{p_{0}}$ this metric corresponds to the one found in 
\cite{Kempf2}, which in that formulation may be obtained through integration
by parts.

Next we select the deformed harmonic oscillator as a specific example
supplemented by the deformed commutation relations (\ref{12}) together with
the above mentioned representations for $X$ and $P$%
\begin{eqnarray}
H_{ho} &=&\frac{P^{2}}{2m}+\frac{m\omega ^{2}}{2}X^{2},  \notag \\
&=&\frac{p_{0}^{2}}{2m}+\frac{m\omega ^{2}}{2}(1+\tau p_{0}^{2})x_{0}(1+\tau
p_{0}^{2})x_{0},  \label{3} \\
&=&\frac{p_{0}^{2}}{2m}+\frac{m\omega ^{2}}{2}\left[ (1+\tau
p_{0}^{2})^{2}x_{0}^{2}+2i\hbar \tau p_{0}(1+\tau p_{0}^{2})x_{0}\right] . 
\notag
\end{eqnarray}
As the dynamical variables $X$ and $P$ are no longer Hermitian, the standard
harmonic oscillator Hamiltonian in terms of the variables $x_{0}$ and $p_{0}$
ceases to be Hermitian as well when replacing $(x_{0}$,$p_{0})$ by $(X$,$P)$ 
\begin{equation}
H_{ho}^{\dagger }(X,P)=\rho (X,P)H_{ho}(X,P)\rho ^{-1}(X,P).  \label{212}
\end{equation}
By construction the Hamiltonian in (\ref{3}) is pseudo and quasi Hermitian
as can be checked very easily. Moreover the third version in (\ref{3}) could
be a standard starting point in the context of non-Hermitian $\mathcal{PT}$%
-symmetric quantum mechanical models, as the Hamiltonian evidently respects
this symmetry, $\left[ \mathcal{PT}\text{,}H_{ho}\right] =0$. The
simultaneous parity transformation $\mathcal{P}$ and time reversal $\mathcal{%
T}$ are realized as $\mathcal{P}:x_{0}\rightarrow -x_{0}$, $p_{0}\rightarrow
-p_{0}$; $\mathcal{T}:x_{0}\rightarrow x_{0}$, $p_{0}\rightarrow -p_{0}$, $%
i\rightarrow -i$. The undeformed and deformed dynamical variables transform
in the same manner under a $\mathcal{PT}$-operation, such that the $\mathcal{%
PT}$-symmetry is preserved in the deformation process $H(x_{0},p_{0})%
\rightarrow H(X,P)$.

Next we compute the corresponding quantities in the standard framework of
the Hermitian counterpart. From the explicit form of $\rho $ it is trivial
to find the Dyson map $\eta =(1+\tau P^{2})^{-1/2}$ according to (\ref{sum}%
), such that the physical variables corresponding to momentum and position
in the Hermitian system result according to (\ref{4}) to 
\begin{equation}
x=(1+\tau p_{0}^{2})^{1/2}x_{0}(1+\tau p_{0}^{2})^{1/2}~~~\text{and~~~}%
~p=p_{0}.  \label{hrep}
\end{equation}
These operators satisfy the same deformed canonical commutation relations as
their counterparts in the non-Hermitian version of the theory. Consequently
the Hermitian counterpart Hamiltonian becomes 
\begin{eqnarray}
h_{ho} &=&\eta H_{ho}\eta ^{-1}=\frac{p^{2}}{2m}+\frac{m\omega ^{2}}{2}x^{2},
\label{hho} \\
&=&\frac{p_{0}^{2}}{2m}\!+\!\frac{m\omega ^{2}}{2}(1+\tau p_{0}^{2})^{\frac{1%
}{2}}x_{0}(1+\tau p_{0}^{2})x_{0}(1+\tau p_{0}^{2})^{\frac{1}{2}}\!,  \notag
\\
&=&\left( \frac{1}{2m}-m\omega ^{2}\hbar ^{2}\tau ^{2}\right) p_{0}^{2}+%
\frac{m\omega ^{2}}{2}\left[ (1+\tau p_{0}^{2})^{2}x_{0}^{2}\right.  \notag
\\
&&\qquad \ \ \qquad \qquad \left. +4i\hbar \tau p_{0}(1+\tau
p_{0}^{2})x_{0}-\hbar ^{2}\tau \right] .  \notag
\end{eqnarray}
Recalling that the commutation relations were the starting point in the
first place, we notice that the non-Hermitian nature of the construction
could have been avoided from the beginning when selecting the equivalent
Hermitian representation (\ref{hrep}) right from the start, although a
priori this would be less obvious to guess.

Taking now $g(N)=N/\gamma $ and parameterizing the deformation parameter in
the form $q=e^{\tau \gamma ^{3}}$ with $\tau \in \mathbb{R}^{+}$, the limit $%
\gamma \rightarrow 0$ yields an exponential deformation of the canonical
commutation relations
\begin{equation}
\left[ X,P\right] =i\hbar e^{\tau P^{2}}.  \label{C2}
\end{equation}
Representations for $X$ and $P$ in terms of $x_{0}$ and $p_{0}$ are easily
found. We choose $X=e^{\tau p_{0}^{2}}x_{0}$ and $P=p_{0}$, which reduces to
the previous deformation to first order in $\tau $. As we found also above,
the representations are not Hermitian $X^{\dagger }=X+2\tau i\hbar Pe^{\tau
P^{2}}~$and$~P^{\dagger }=P$. The unique metric operator which conjugates $X$
and $P$ is then found to be $\rho =e^{-\tau P^{2}}$.

Let us now specify a concrete Hamiltonian depending on these variables. Of
course we could also study the deformed harmonic oscillator in terms of the
variables corresponding to (\ref{C2}) along the lines outlined above, but
instead we investigate a non-Hermitian version of it, the so-called Swanson
model \cite{Swanson}
\begin{equation}
H_{S}(X,P)=\frac{P^{2}}{2m}+\frac{m\omega ^{2}}{2}X^{2}+i\mu \{X,P\}.
\label{swan}
\end{equation}
Even in the limit $\tau \rightarrow 0$ this Hamiltonian remains
non-Hermitian and consequently replacing the non-Hermitian variables $(X,P)$
by their Hermitian counterparts according to (\ref{sum}) 
\begin{equation}
x=e^{\frac{\tau }{2}p_{0}^{2}}x_{0}e^{\frac{\tau }{2}p_{0}^{2}}\quad \quad 
\text{and\quad \quad }p=p_{0},
\end{equation}
with $\eta =e^{-\frac{\tau }{2}P^{2}}$ will not render it Hermitian
\begin{equation}
H_{S}(x,p)=\frac{x^{2}}{2m}+\frac{m\omega ^{2}}{2}x^{2}+i\mu \hbar m\omega
^{2}\{x,p\}.
\end{equation}
Nonetheless, we may invoke the steps indicated in (\ref{HH}) and find an
isospectral Hermitian counterpart 
\begin{equation}
h_{S}(x,p)=\left( \frac{1}{2m}+2\mu ^{2}\hbar ^{2}m\omega ^{2}\right) p^{2}+%
\frac{m\omega ^{2}}{2}x^{2},
\end{equation}
using the transformation $\tilde{\eta}H_{S}(x,p)\tilde{\eta}^{-1}=h_{S}(x,p)$
with $\tilde{\eta}=$ $e^{\mu p_{0}^{2}}$. A crucial point to notice is that
the Hermitian variables $(x,p)$ are only observables in the Hermitian system 
$h_{S}(x,p)$, whereas the counterparts in the deformed non-Hermitian system
are $\tilde{X}=\tilde{\eta}^{-1}\eta X\eta ^{-1}\tilde{\eta}~$and$~\tilde{P}=%
\tilde{\eta}^{-1}\eta P\eta ^{-1}\tilde{\eta}=p_{0}.$

Similar arguments will always hold when $\tilde{\eta}$ and $\eta $ commute,
which for the deformations (\ref{12}) and (\ref{C2}) is the case when $%
\tilde{\eta}$ only depends on\ $p_{0}$. A further non-trivial example for
this is for instance the $-x^{4}$-potential, which despite being unbounded
from below can be shown to posses a well defined spectrum when making a
suitable variable transformation and a subsequent similarity transformation
of the form $\tilde{\eta}=$ $e^{\alpha p_{0}^{3}+\beta p_{0}}$ with $\alpha
,\beta \in \mathbb{R}$ \cite{JM}.

\section{Minimal length}

An important physical consequence resulting from the deformation of the
canonical commutation relations is the unavoidable occurrence of a minimal
length. Considering the uncertainty relations for the deformed canonical
observables of the system (\ref{swan})
\begin{equation}
\Delta \tilde{X}\Delta \tilde{P}=\Delta x\Delta p\geq \frac{1}{2}\left|
\left\langle [\tilde{X},\tilde{P}]\right\rangle _{\tilde{\rho}\rho
^{-1}}\right| =\frac{1}{2}\left| \left\langle [x,p]\right\rangle \right| 
\label{HU}
\end{equation}
it is clear that a minimal length must always arise once the right hand side
of (\ref{HU}) involves higher powers of $\tilde{P}$. One is then naturally
led to a minimal uncertainty in the limit $\Delta \tilde{X}\rightarrow 0$ as
then the momentum uncertainty $\Delta \tilde{P}$ becomes very large, but
eventually the linear behaviour on the left hand side of the inequality will
be too weak to balance the higher powers of $\tilde{P}$ on the right hand
side. Consequently the limit $\Delta \tilde{X}\rightarrow 0$ can not be
reached without violating the inequality (\ref{HU}) and a localization in
space is no longer possible.

We now compute the explicit value for the minimal length for the deformed
commutation relation (\ref{C2}) associated to the Hamiltonian (\ref{swan}) 
\begin{equation}
\Delta \tilde{X}\Delta \tilde{P}\geq \frac{i\hbar }{2}e^{\tau \left\langle 
\tilde{P}^{2}\right\rangle _{\tilde{\rho}\rho ^{-1}}}=\frac{i\hbar }{2}%
e^{\tau \left[ \left\langle \tilde{P}\right\rangle _{\tilde{\rho}\rho
^{-1}}^{2}+(\Delta \tilde{P})^{2}\right] }.
\end{equation}
In order to determine the minimal value for $\Delta \tilde{X}$ we have to
solve the two equations
\begin{equation}
\partial _{\Delta \tilde{P}}f(\Delta \tilde{X},\Delta \tilde{P})=0\quad
\quad \text{and\quad \quad }f(\Delta \tilde{X},\Delta \tilde{P})=0,
\label{f2}
\end{equation}
with $f(\Delta \tilde{X},\Delta \tilde{P})=\Delta \tilde{X},\Delta \tilde{P}-%
\frac{i\hbar }{2}e^{\tau \left[ \left\langle \tilde{P}\right\rangle _{\tilde{%
\rho}\rho ^{-1}}^{2}+(\Delta \tilde{P})^{2}\right] }$ for $\Delta \tilde{X}$%
. There is no general solution to this equation, but we may find a minimal
value order by order in $\tau $, when expanding $f(\Delta \tilde{X},\Delta 
\tilde{P})$ in powers of $\tau $ as $f(\Delta \tilde{X},\Delta \tilde{P}%
)=a_{0}+a_{1}\tau +a_{2}\tau ^{2}+\ldots $ To first order we find the
minimal uncertainty
\begin{equation}
\Delta \tilde{X}_{\min }^{(1)}=\hbar \sqrt{\tau }\sqrt{1+\tau \left\langle 
\tilde{P}\right\rangle _{\tilde{\rho}\rho ^{-1}}^{2}},
\end{equation}
which corresponds to the value already found in \cite{Kempf2}. This is to be
expected as to that order the deformed commutation relations (\ref{12}) and (%
\ref{C2}) coincide. Expanding $f$ to second order and solving (\ref{f2})
thereafter yields
\begin{eqnarray}
(\Delta \tilde{X}_{\min }^{(2)})^{2} &=&\frac{\hbar ^{2}\tau }{27}\left\{ %
\left[ 7+4\tau \left\langle p\right\rangle ^{2}(2+\tau \left\langle
p\right\rangle ^{2})\right] ^{\frac{3}{2}}\right.  \\
&&\left. +(1+\tau \left\langle p\right\rangle ^{2})\left[ 17+8\tau
\left\langle p\right\rangle ^{2}(2+\tau \left\langle p\right\rangle ^{2})%
\right] \right\} ,  \notag
\end{eqnarray}
where we used $\left\langle \tilde{P}\right\rangle _{\tilde{\rho}\rho
^{-1}}^{2}=\left\langle p\right\rangle $. Since $\tau $ is positive the
smallest values this expression can acquire, say $\Delta \tilde{X}%
_{0}^{(\kappa )}$ with $\kappa $ denoting the order, occur when $%
\left\langle p\right\rangle =0$ since $\left\langle p\right\rangle $ is
always real. We perform this minimization order by order in units of $\hbar 
\sqrt{\tau }$: $\Delta \tilde{X}_{0}^{(1)}=1$, $\Delta \tilde{X}_{0}^{(2)}=%
\sqrt{\frac{17+7\sqrt{7}}{27}}~=1.14698088$, $\Delta \tilde{X}%
_{0}^{(3)}=1.16373131$, $\Delta \tilde{X}_{0}^{(4)}=1.16562060$, $\Delta 
\tilde{X}_{0}^{(5)}=1.16580546$, $\Delta \tilde{X}_{0}^{(6)}=1.16582082$, $%
\Delta \tilde{X}_{0}^{(7)}=1.16582191$, $\Delta \tilde{X}%
_{0}^{(8)}=1.16582198$, $\Delta \tilde{X}_{0}^{(9)}=1.1658219905$, $\Delta 
\tilde{X}_{0}^{(10)}=1.1658219907$. Thus for the deformed commutation
relations (\ref{C2}) we observe a fast convergence of the absolute minimal
length to a value of $\Delta \tilde{X}_{0}\approx 1.16582199\hbar \sqrt{\tau 
}$. This means the minimal length resulting either from the deformation (\ref
{12}) or (\ref{C2}) differ by around $16\%$, which allows for new
opportunities in the context of theories describing quantum gravity.

\section{Conclusions}

We demonstrated that non-Hermitian Hamiltonian systems and deformed quantum
mechanics based on deformations of the uncertainty relations may be treated
in a similar fashion. The key difference between the two scenarios is that
in the former the starting point are Hamiltonians whereas in the latter one
commences with a set of dynamical variables. Uniqueness of the construction
is therefore only guaranteed in the latter case. The absolute minimal length
is entirely governed by the choice of the observables and not by the
explicit form of the underlying Hamiltonian. Representation for the generic
version (\ref{gxp}) will lead to more involved non-Hermitian Hamiltonians,
different minimal uncertainties and lengths and in addition to minimal
momenta.

\textbf{Acknowledgments}: For one of us (BB), it is a great pleasure to
thank Prof Andreas Fring, as well as other members of the Centre for
Mathematical Science at City University for providing a perfect ambience for
research.


\end{document}